\documentclass[11pt,twoside]{article}
\usepackage{asp2010}
\usepackage{psfig,epsfig}

\resetcounters

\begin{document}

\nocite{2003A&A...412..813R}

\title{Rotation and differential rotation in rapidly rotating field stars}
\author{Matthias~Ammler-von Eiff, Ansgar~Reiners
\affil{Institut f\"ur Astrophysik, Georg-August-Universit\"at, Friedrich-Hund-Platz 1, 37077 G\"ottingen, Germany}}

\begin{abstract}
We continue our studies on stellar latitudinal differential rotation. The presented work is a sequel of the work of Reiners et al. who studied the spectral line broadening profile of hundreds of stars of spectral types A through G at high rotational speed ($v\sin{i}\,>\,12\,$km\,s$^\mathrm{-1}$). While most stars were found to be rigid rotators, only a few tens show the signatures of differential rotation.
The present work comprises the rotational study of some 180 additional stars. The overall broadening profile is derived according to Reiners et al. from hundreds of spectral lines by least-squares deconvolution, reducing spectral noise to a minimum. Projected rotational velocities $v\,\sin{i}$ are measured for about 120 of the sample stars. Differential rotation produces a cuspy line shape which is best measured in inverse wavelength space by the first two zeros of its Fourier transform. Rigid and differential rotation can be distinguished for more than 50 rapid rotators ($v\sin{i}\,>\,12\,$km\,s$^\mathrm{-1}$) among the sample stars from the available spectra.
Ten stars with significant differential rotation rates of 10-54\,\% are identified, which add to the few known rapid differential rotators. Differential rotation measurements of 6\,\% and less for four of our targets are probably spurious and below the detection limit. Including these objects, the line shapes of more than 40 stars are consistent with rigid rotation.
\end{abstract}

\section{Introduction}
Radial and latitudinal differential rotation on the Sun is thought to be formed by the interaction of rotation and the convective envelope. Differential rotation is a main ingredient of the solar dynamo. Latitudinal differential rotation is described by a simple surface rotation law with dependence on the latitude $l$ and the parameter of differential rotation $\alpha$:
￼\begin{equation}
\Omega(l)=\Omega_\mathrm{Equator}(1-\alpha\sin^2{l})
\end{equation}
Differential rotation of stars has been studied extensively in recent years by Doppler imaging \citep{2005MNRAS.357L...1B} and line profile analysis \citep{2006A&A...446..267R}.

\section{Sample}
\begin{figure}[ht!]
\plotone{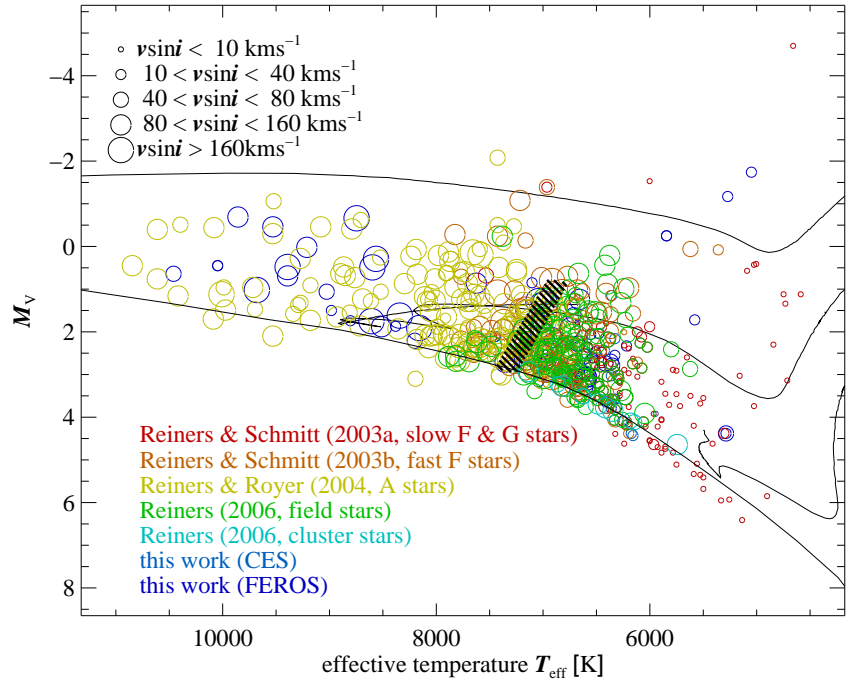}
\caption{\label{fig:hrd_vsini}The HR diagram of all stars analyzed by line profile analysis including the sample of the present work. Symbol size scales with projected rotational velocity $v\,\sin{i}$ as indicated in the figure. Solid lines indicate evolutional tracks for 1, 2, and 5\,$M_\odot$ with overshooting and the early main-sequence according to \citet{SDF00}. The hatched area shows the approximate location of the granulation boundary \citep{1989ApJ...341..421G}.}
\end{figure}

The present work adds 184 stars of the southern sky known to rotate faster than about 10\,km\,s$^{-1}$, to be brighter than $V=6$, and with colours $0.3\,<\,(B-V)\,<\,0.9$. Emphasis was given to the selection of suitable targets rather than achieving a complete or unbiased sample. Figure~\ref{fig:hrd_vsini} shows the HR diagram of all stars analyzed in previous work and the present work by line profile analysis including models by \citet{SDF00} and the granulation boundary according to \citet{1989ApJ...341..421G}. Cool stars are affected by magnetic braking. Therefore, rotational speed is correlated with effective temperature, so that these two parameters are degenerate. It is challenging to disentangle the effects of effective temperature and rotational speed onto differential rotation \citep{2006A&A...446..267R}.

\section{Observations}
Spectra with high signal-to-noise ratio and a resolving power of 48,000 were taken with FEROS at the ESO/MPG-2.2m telescope, La Silla (Chile), covering the wavelengths between 360\,nm and 920\,nm. Spectra of 24 stars with projected rotational velocities of less than 45\,km\,s$^\mathrm{-1}$ were obtained with the CES spectrograph at the ESO-3.6m telescope, La Silla, at a spectral resolving power of 220,000 and a wavelength range of 614\,nm - 617.5\,nm.

\section{Line profile analysis}
\begin{figure}[ht!]
\plotone[width=\textwidth]{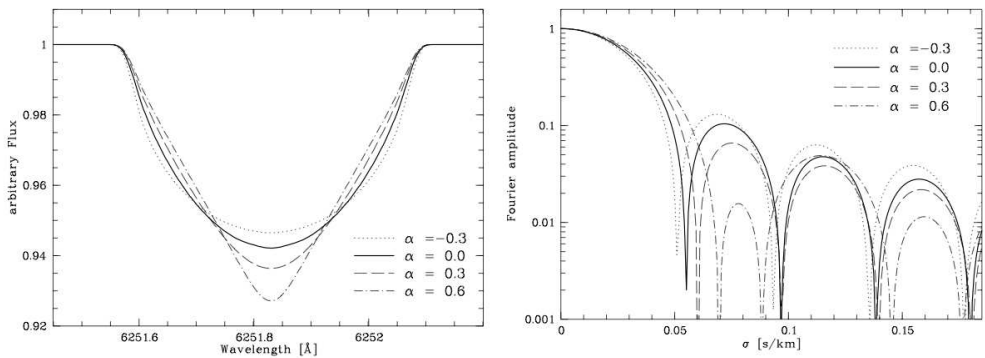}
\caption{\label{fig:Reiners02_3}Models of broadening profiles for different amounts of differential rotation $\alpha$ in wavelength space (left) and Fourier space (right). The location of the first two zeros of the Fourier transform $q_1$ and $q_2$ is used to measure the amount of differential rotation. Credit: \citet{2002A&A...384..155R} reproduced with permission \copyright~ESO.}
\end{figure}

An average line profile is derived by least-squares deconvolution of a wide spectral range containing hundreds of spectral lines \citep{2002A&A...384..155R}. The line shape is best measured in Fourier space by the ratio of the first two zeros of the Fourier transform $\frac{q_\mathrm{2}}{q_\mathrm{1}}$ (Fig.~\ref{fig:Reiners02_3}). This ratio is calibrated in terms of the parameter of differential rotation $\alpha$ \citep{2003A&A...398..647R}. In the case of a rigid rotator, $\frac{q_\mathrm{2}}{q_\mathrm{1}}$ varies between 1.72 and 1.83 depending on the limb darkening. Lower values indicate differential rotation while larger values represent the rather hypothetical case of anti-solar differential rotation which is most probably caused by a polar spot in the regime of rigid rotation. The derivation of the rotational profile is restricted to stars with a projected rotational velocity $v\,\sin{i}$ of at least 10\,km\,s$^\mathrm{-1}$. Thus, most cool dwarf stars and giants cannot be analyzed.

\section{Results}
\begin{figure}[ht!]
\plotone{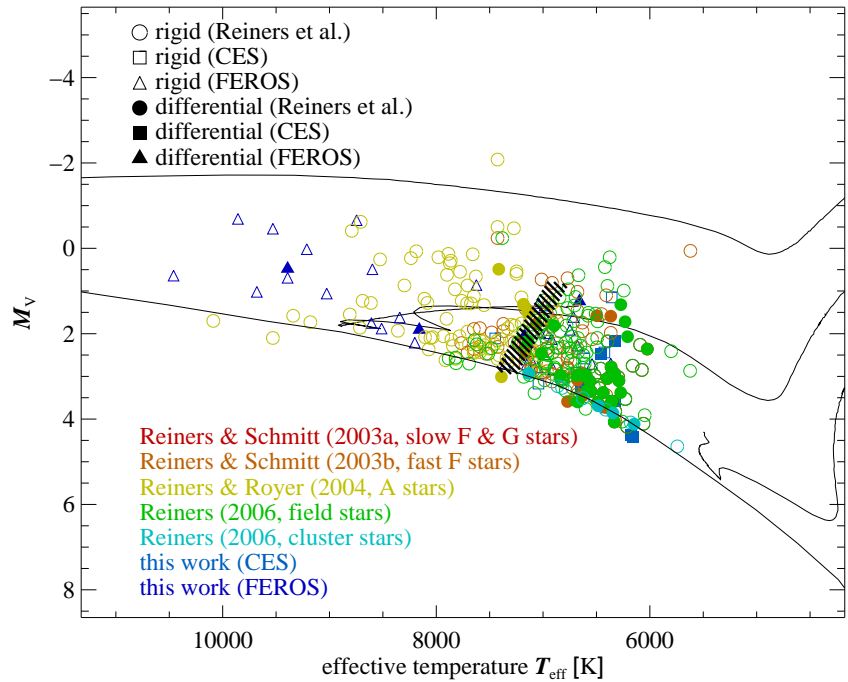}
\caption{\label{fig:hrd_diffrot}The HR diagram of Fig.~\ref{fig:hrd_vsini}, now with symbols indicating the measured mode of rotation.}
\end{figure}

\begin{figure}[ht!]
\plotone[width=10cm]{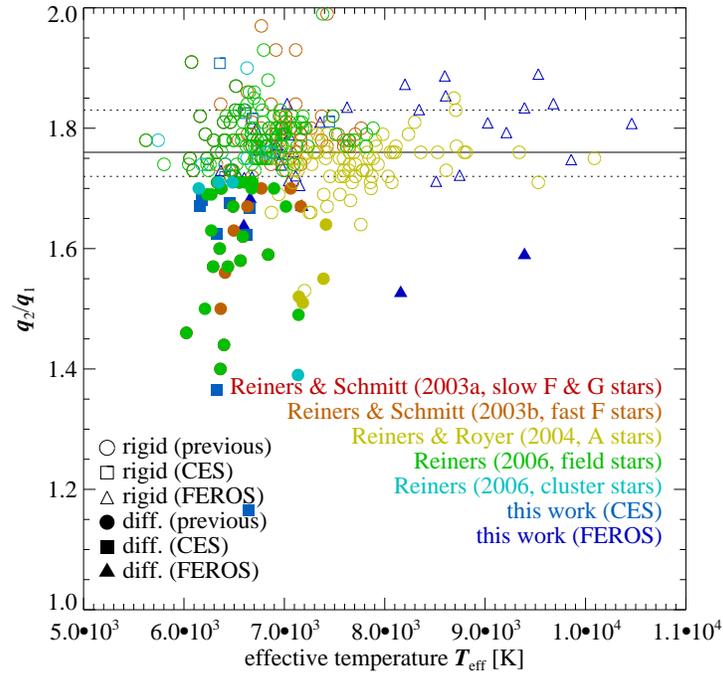}
\caption{\label{fig:comp_tdiffrot}The ratio of the zeros of the Fourier transform of the broadening profile $\frac{q_\mathrm{2}}{q_\mathrm{1}}$ which indicates the amount of differential rotation is plotted vs. effective temperature. The solid line marks the position of rigid rotators for solar limb darkening with a limb darkening parameter of $\epsilon = 0.6$, the region between the dotted lines for any value of $\epsilon$ between 0.0 and 1.0. Full symbols below these lines indicate probable differential rotators while the other symbols in this region represent probable spurious measurements of differential rotation.}
\end{figure}

\begin{figure}[ht!]
\plotone[width=10cm]{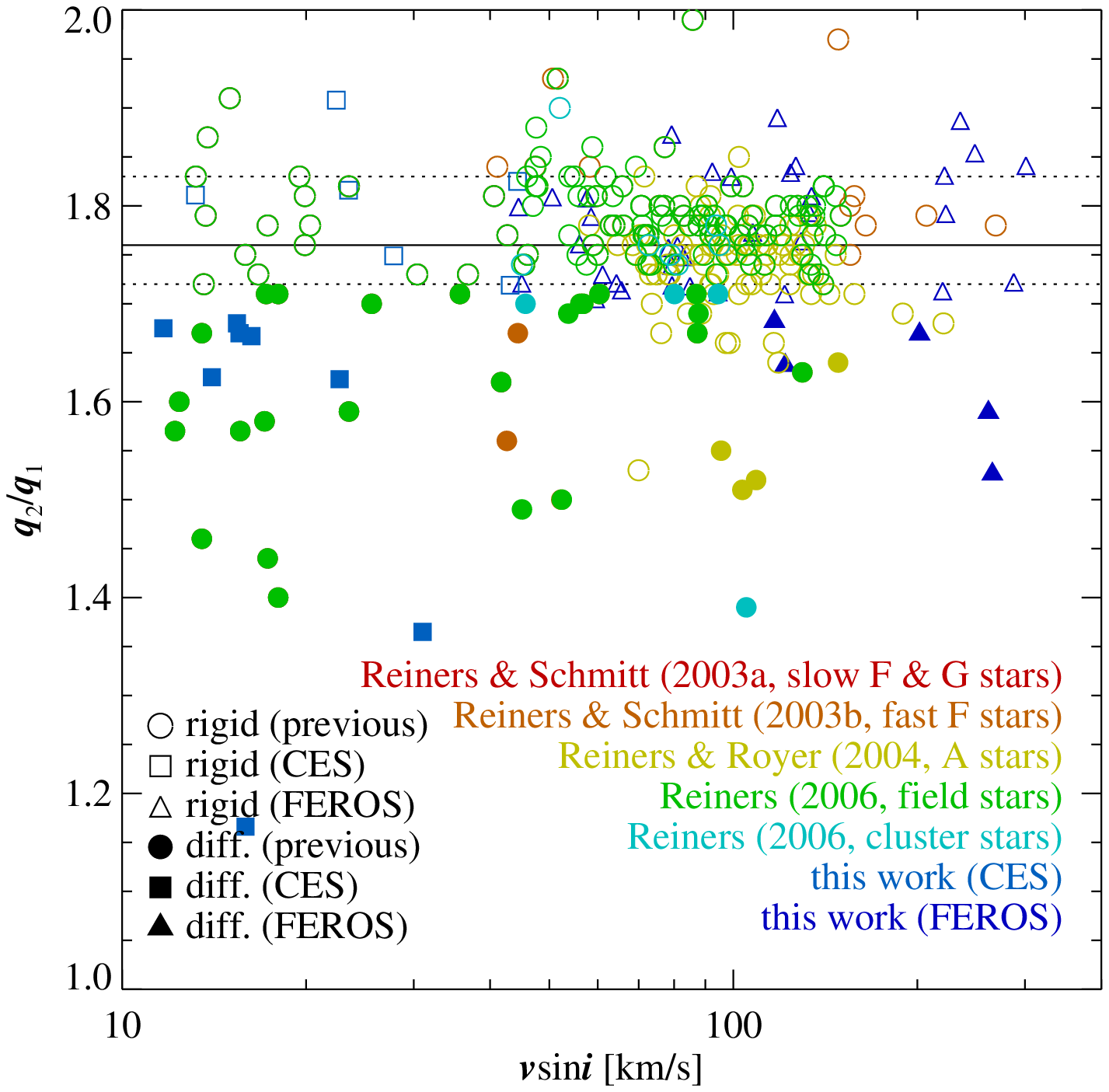}
\caption{\label{fig:comp_vdiffrot}Similar to Fig.~\ref{fig:comp_tdiffrot}, $\frac{q_\mathrm{2}}{q_\mathrm{1}}$ is plotted vs. projected rotational velocity $v\,\sin{i}$.}
\end{figure}

The HR diagram (Fig.~\ref{fig:hrd_vsini}) is repeated in Fig.~\ref{fig:hrd_diffrot} now showing the distribution of differential and rigid rotators. The correlation of the amount of differential rotation indicated by the ratio $\frac{q_\mathrm{2}}{q_\mathrm{1}}$ with effective temperature and projected rotational velocity is given in Figs.~\ref{fig:comp_tdiffrot} and \ref{fig:comp_vdiffrot}, respectively. Ten stars show clear signatures of differential rotation between 10\,\% and 54\,\%. The amount of differential rotation of four stars is below 6\,\% which is below our detection limit and consistent with rigid rotation. Two F stars (HD\,104731 and HD\,124425) with $T_\mathrm{eff}\approx6500\,$K and $v\,\sin{i}\lesssim30\,$km\,s$^\mathrm{-1}$ display the lowest values of $\frac{q_\mathrm{2}}{q_\mathrm{1}}$ measured so far which correspond to a differential rotation of 54\,\% and 44\,\%, respectively.  At the hot side of the granulation boundary, the frequency of differential rotators diminishes, i.e. $\frac{q_\mathrm{2}}{q_\mathrm{1}}$ has values not much less than 1.72-1.83 (Fig.~\ref{fig:comp_tdiffrot}). Three new candidates rotate faster than 200\,km\,s$^\mathrm{-1}$. In these cases, the effect might be due to gravitational darkening in the regime of rigid rotation \citep{2003A&A...408..707R,2004A&A...415..325R}. The spherical shape of the stellar surface is distorted by centrifugal forces and the resulting temperature variations modify the surface flux distribution. Five stars show signatures which can be explained by anti-solar differential rotation or more plausibly by rigid rotation with polar spots. In total, there are 44 stars with spectra consistent with rigid rotation.

\acknowledgements 
M.A. and A.R. acknowledge research funding granted by the Deutsche Forschungsgemeinschaft (DFG) under the project RE 1664/4-1. This research has made use of the extract stellar request type of the Vienna Atomic Line Database (VALD), the Simbad and VizieR databases, operated at CDS, Strasbourg, France, and NASA's Astrophysics Data System Bibliographic Services.

\bibliography{ammler_m}

\end{document}